\documentclass[preprintnumbers,showpacs,amsmath,amssymb,twocolumn,floatfix,aps,prd,nofootinbib,showkeys]{revtex4-1}
\usepackage{graphicx}
\usepackage{bbm}
\usepackage{bm}
\usepackage{amsfonts}
\usepackage{epstopdf}
\usepackage[colorlinks,linkcolor=red,anchorcolor=blue,citecolor=green]{hyperref}
\usepackage{subfigure}
\usepackage{float}
\usepackage{enumerate}
\newcommand\diff{\mathrm{d}}

\begin{document}
\title{
	New proper tetrad for teleparallel gravities in non-flat spacetime
}
\author{Rui-Hui \surname{Lin}}
\email[]{linrh@shnu.edu.cn}
\author{Xiang-Hua \surname{Zhai}}
\email[]{zhaixh@shnu.edu.cn}
\affiliation{Shanghai United Center for Astrophysics (SUCA), Shanghai Normal University,
100 Guilin Road, Shanghai 200234, China}

\begin{abstract}
The restoration of spin connection clarifies the long known local Lorentz invariance problem in telelparallel gravities.
It is considered now that any tetrad together with the associated spin connection can be equally utilized.
Among the tetrads there is a particular one, namely proper tetrad,
in which all the spurious inertial effects are removed and the spin connection vanishes.
A specific tetrad was proposed in the literature for spherically symmetric cases,
which has been used in regularizing the action, as well as in searching solutions in various scenarios.
We show in this paper that the this tetrad is not the unique choice for the proper tetrad.
We construct a new tetrad that can be considered as the proper one,
and it will lead to different behaviors of the field equation and results in different solutions.
With this proper tetrad,
it is possible to find solutions to teleparallel gravities in the strong field regime, which may have physical applications.
In the flat spacetime limit, the new tetrad coincides with the aforementioned one.
\end{abstract}
\maketitle

\section{Introduction}
General Relativity (GR) seems to work perfectly well
against the local weak field tests of gravity,
whereas the long known challenge of it when applied to the entire Universe,
i.e. the dark contents of the Universe,
still lacks a consensus solution,
which motivates the substantial study of alternative theories of gravity.
The recent detection of
gravitational waves from
binary black holes and neutron stars
has now opened a new window to test these theories in the strong field regime.
Most of the study regarding the modified gravities in the literature
are based on the curvature formulation.
Besides that, one can also search for alternative theories
starting from the Teleparallel Equivalent of General Relativity (TEGR)
\cite{aldrovandi2012teleparallel,Maluf2013},
in which the tetrad field is used as the dynamical variable instead of the metric,
and torsion scalar $T$ is constructed to be the underlying Lagrangian.
The simplest modifying scheme under this framework is the $f(T)$ theories
\cite{Bengochea2009,Linder2010,Cai2016}.
When non-linear Lagrangian is constructed,
the field equations in $f(T)$ models feature in having second order derivatives as in GR
instead of fourth order ones in $f(R)$ models.
Simpler may be the field equation, however,
the strong field solutions in teleparallel gravities are still difficult to extract.
One of the difficulties lies in the misunderstanding about their lack of local Lorentz invariance.
It is not new that the torsion-based theories suffer the problem of local covariance
\cite{Sotiriou2011,Li2011,Ferraro2015,Bahamonde2015},
which comes from the simply setting of vanishing spin connection.
The frame-dependent nature of this pure-tetrad formalism
has led to efforts in finding the ``good'' tetrad in certain situations
\cite{FERRARO2011,Tamanini2012},
but the head-on solution to this problem should be restoring the spin connection
and constructing the covariant formulation of teleparallel gravities
\cite{Obukhov2006,Lucas2009,Krssak2016,Golovnev2017,Krssak2018}.
There is no longer a privileged ``good'' tetrad in the sense that
any tetrad within a  Lorentz group can be used as long as the associated spin connection is restored.
Nonetheless, the determination of this spin connection,
or equivalently the determination of the proper tetrad,
is still open to debate.
The theory will take different form with different choice of proper tetrad.
A specific tetrad was proposed in Refs.\cite{Lucas2009,Krssak2016} to be the proper one for spherically symmetric cases,
which has been used in the study of the covariant teleparallel gravities in various scenarios
\cite{Krssak2016,Ruggiero2015,DeBenedictis2016,Farrugia2016,Lin2017,Mai2017,Pace2017}.
In this paper,
we show that this tetrad is not the unique choice for the proper tetrad.
We construct a new proper tetrad which will lead to a different description of the theory
and may be used to search for strong field solutions in such spacetime.
In the flat spactime limit,
the two tetrads coincide with each other.

The paper is organized as follows.
We briefly review the covariant teleparallel gravities in Section \ref{review}.
In Sec. \ref{newtetrad}, we construct a new tetrad that can be considered as the proper one for static spherically symmetric metric.
The flat limit of the new proper tetrad and some examples are presented in Sec. \ref{flatlimit}.
Sec. \ref{discussions} contains our conclusion and discussions.

\section{Tetrad and spin connection}
\label{review}
As the spacetime manifold $\mathcal M$ is assumed to be a metric space with $g$ and parallelizable,
it is generally possible to find a trivialization $e_a=e_a^{\:\mu}\partial_\mu$ of the tangent bundle of the manifold.
The dual vector basis 1-form to $e_a$, i.e. the tetrad, is given by $h^a=h^a_{\:\mu}\diff x^\mu$.
The line element is then written as $\diff s^2=g_{\alpha\beta}\diff x^\alpha\otimes\diff x^\beta=\eta_{ab}h^a\otimes h^b$,
where $\eta_{ab}$ is the Minkowski metric of the tangent space.
The torsion 2-form is given by\cite{aldrovandi2012teleparallel,Krssak2016}
\begin{equation}
	T^a=\mathcal D h^a=\diff h^a+\omega^a_{\:b}\wedge h^b,
	\label{covariantT}
\end{equation}
where the covariant exterior derivative $\mathcal D$ and the spin connection $\omega^a_{\:b}$ are introduced
such that for any vector $V^a$ in the tangent space at a given point,
$\mathcal D_\mu V^a$ is covariant under Lorentz rotation.
That is, for a rotation of frame $h^a\mapsto\mathring h^a=\Lambda^a_bh^b$ with $\Lambda^a_b(x)\in\text{SO}(1,3)$,
the derivative transforms into
$\mathring{\mathcal D}\mathring V^a=\Lambda^a_b\mathcal D V^b$.
This ensures the torsion 2-form \eqref{covariantT} is locally Lorentz invariant,
and requires the spin connection to satisfy
\begin{equation}
	\omega^a_{\;d}=(\Lambda^{-1})^a_b\mathring\omega^b_{\;c}\Lambda_d^c+(\Lambda^{-1})_c^a(\diff\Lambda^c_d).
	\label{omegatrans}
\end{equation}
Generally,
one can always find a specific frame in which all components of the spin connection vanish.
Suppose we have found such a frame $\mathring h^b$ and it connects to the original frame $h^a$ via the transform $\Lambda^b_a$.
Then $\mathring\omega^b_{\;c}=0$ and
\begin{equation}
	\omega^a_{\;b}=(\Lambda^{-1})_c^a(\diff\Lambda^c_b),
	\label{omega0}
\end{equation}
i.e. purely determined by the Lorentz transform.
The tetrad with vanishing spin connection is then called the proper tetrad\cite{Lucas2009}.
And the spin connection reflects the spurious inertial effects
of the given frame with respect to the proper frame.
Reversely, to determine the spin connection is in fact to determine $\Lambda$ via Eq.\eqref{omega0},
which will transform the given tetrad into the proper one.

With these setup, the torsion scalar is then given by
\begin{equation}
		T=T^a\wedge\star\left( T_a-h^a\wedge\left( e_b\cdot T^b \right)-\frac12 e^a\cdot\left( h^b\wedge T_b \right) \right)
	\label{exttorsion}
\end{equation}
where $\star$ denotes the Hodge dual and $\cdot$ indicates the interior product.
TEGR takes this scalar as its Lagrangian,
and the $f(T)$ gravities, on the other hand, consider an arbitrary function of $T$ instead.

Historically, the torsion 2-form was defined using the simple exterior derivative with vanishing spin connection
in the so-called pure-tetrad formalism.
This made the torsion 2-form not covariant under local Lorentz transform .
In the case of TEGR, the difference lies in the surface term and hence not much problem arises.
The non-trivial $f(T)$ gravities, however, take different forms depending on the tetrad in this pure-tetrad formalism,
and some choice of tetrad is considered ``better'' than others.
The covariant derivative introduced in \eqref{covariantT} deprives the privileges of any ``good'' tetrad
and clarifies the misunderstanding about the lack of local Lorentz invariance of the teleparallel gravities.
Nonetheless, the determination of the proper tetrad is still the crux of the whole covariant construction.

\section{A new tetrad for non-flat spacetime}
\label{newtetrad}
As mentioned before,
a given metric and a specific tetrad along with the determination of the spin connection in fact establishes a tetrad that considered to be proper.
For example, in the static spherically symmetric case
\begin{equation}
	\diff s^2=\xi\diff t^2-\zeta\diff r^2-r^2\diff\Omega^2
	\label{sssmetric}
\end{equation}
with the diagonal tetrad
\begin{equation}
	\begin{split}
	&h^{\hat0}=\sqrt\xi\diff t,\quad h^{\hat1}=\sqrt\zeta\diff r,\\
	&h^{\hat2}=r\diff\theta,\quad h^{\hat3}=r\sin\theta\diff\phi,
\end{split}
	\label{diage}
\end{equation}
where $\xi(r),\zeta(r)$ are functions of the radial coordinate $r$,
the determination of the spin connection gives, via Eq.\eqref{omega0}, a Lorentz rotation
that transforms the diagonal tetrad \eqref{diage} into the proper tetrad.
Thus we search for the appropriate transform $\Lambda$ directly.
In general, a Lorentz transform $\Lambda\in\text{SO}(1,3)$ can be decomposed into
\begin{equation}
	\Lambda=\mathcal R_3\mathcal B,
	\label{lambda0}
\end{equation}
where $\mathcal R_3\in\text{SO}(3)$ is the spatial rotation,
and $\mathcal B$ is a Lorentz boost.

Since it is the proper frame that we are seeking,
we \emph{assume} that the Lorentz boost should be such that
it takes the frame into a freely falling one.
Thus,
\begin{equation}
	\mathcal B=\left(
	\begin{array}{cccc}
		\frac1{\sqrt\xi}&-\sqrt{\frac{1-\xi}\xi}&0&0\\
		-\sqrt{\frac{1-\xi}\xi}&\frac1{\sqrt\xi}&0&0\\
		0&0&1&0\\
		0&0&0&1
	\end{array}
	\right)
	\label{boost}
\end{equation}
along $\hat r$ is proposed.
A similar boost has been considered in Ref.\cite{Ferraro2011a}.

The rotation $\mathcal R_3$ takes place on a spatial slice.
For a fixed time $\diff t=0$, a spatial slice boosted by Eq.\eqref{boost} is given by
$\diff s_{(3)}^2=(\zeta/\xi)\diff r^2+r^2\diff\Omega^2$,
the three legs of the boosted tetrad are
\begin{equation}
	h^{\hat1}_{(3)}=\sqrt{\frac\zeta\xi}\diff r,\quad h^{\hat2}_{(3)}=r\diff\theta,\quad h^{\hat3}_{(3)}=r\sin\theta\diff\phi.
	\label{h3}
\end{equation}
Due the spherical coordinate system,
the pointing directions of these legs are dependent on the angular position of the tetrad,
which obviously brings in spurious inertial effect owing to purely the choice of frame.
So the main idea of the search of the rotation $\mathcal R_3$
is such that it should align the frame everywhere.
However, the underlying spacetime manifold is not flat,
hence the mere alignment may not be enough and the non-flatness should also be accounted for.

In embedding point of view,
$\diff r$ is the projection of real radial increment $\diff\rho$
on the flat slice $\diff r=\cos\alpha\diff\rho$ with $\cos\alpha=\sqrt{\xi/\zeta}$.
With this in mind,
we build a universal Cartesian system
\begin{equation}
	x_{\hat1}=\rho\sin\theta\cos\phi,\quad x_{\hat2}=\rho\sin\theta\sin\phi,\quad x_{\hat3}=\rho\cos\theta
	\label{cartesian}
\end{equation}
to help with the alignment of the tetrads at different position.

Then the angular increment $r\diff\Omega$ can be viewed as one
that deviates from the slice spanned by this system by an angle $\alpha$.
And hence $\diff s_{(3)}^2$ is written as
\begin{equation}
	\diff s_{(3)}^2=\diff\rho^2+\cos^2\alpha(r^2\diff\Omega^2)+\sin^2\alpha(r^2\diff\Omega^2),
	\label{3elem2}
\end{equation}
where the angular increment is divided into two orthogonal parts:
one along the usual angular direction in the Cartesian system given by Eq.\eqref{cartesian},
and the other describes the deviation from it.
One then can decompose the radial increment $\diff\rho$ and the former part of the angular increment $r\diff\Omega$ in the Cartesian system \eqref{cartesian} in the usual way:
\begin{equation}
	\left(
	\begin{array}{c}
		\diff\rho_{\hat1}\\
		\diff\rho_{\hat2}\\
		\diff\rho_{\hat3}
	\end{array}
	\right)=\left(
	\begin{array}{c}
		\sin\theta\cos\phi\diff\rho\\
		\sin\theta\sin\phi\diff\rho\\
		\cos\theta\diff\rho
	\end{array}
	\right),
	\label{rho1stpart}
\end{equation}
and
\begin{equation}
	\left(
	\begin{array}{c}
		r\diff\Omega_{\hat1}\\
		r\diff\Omega_{\hat2}\\
		r\diff\Omega_{\hat3}
	\end{array}
	\right)=\left(
	\begin{array}{cc}
		\cos\theta\cos\phi&-\sin\phi\\
		\cos\theta\sin\phi&\cos\phi\\
		-\sin\theta&0
	\end{array}
	\right)\left(
	\begin{array}{c}
		r\diff\theta\\
		r\sin\theta\diff\phi
	\end{array}
	\right),
	\label{ang1stpart}
\end{equation}
where $\diff\rho_a$ and $r\diff\Omega_a$ are the components of the radial increment and the former part of the angular increment
in the Cartesian system, respectively.

Since the decomposition of the angular increment in Eq.\eqref{3elem2} is orthogonal,
the deviation part should be considered in the perpendicular direction of the former part
(still in the $(\diff\theta,\diff\phi)$ plane though), which, in the Cartesian system, are explicitly:
\begin{equation}
	\left(
	\begin{array}{c}
		r\diff\tilde\Omega_{\hat1}\\
		r\diff\tilde\Omega_{\hat2}\\
		r\diff\tilde\Omega_{\hat3}
	\end{array}
	\right)=\left(
	\begin{array}{cc}
		\pm\sin\phi&\pm\cos\theta\cos\phi\\
		\mp\cos\phi&\pm\cos\theta\sin\phi\\
		0&\mp\sin\theta
	\end{array}
	\right)\left(
	\begin{array}{c}
		r\diff\theta\\
		r\sin\theta\diff\phi
	\end{array}
	\right),
	\label{ang2ndpart}
\end{equation}
where the tilde signs on top of $\diff\Omega$ indicate that they are of the deviation part.

These components in the Cartesian system prescribe naturally the three spatial legs of a tetrad
\begin{equation}
	\mathring{h}^a_{(3)}=\diff\rho_a+\cos\alpha\left( r\diff\Omega_a \right)+\sin\alpha\left( r\diff\tilde\Omega_a \right),
	\label{propertetrad1}
\end{equation}
such that $\diff s_{(3)}^2=\delta_{ab}\mathring h^a\otimes\mathring h^b$.
The transform between this tetrad and the diagonal one \eqref{h3} is
\begin{widetext}
	\begin{equation}
		\mathcal R_3=\left(
		\begin{array}{cccc}
			1&0&0&0\\
			0&\sin\theta\cos\phi&\cos\alpha\cos\theta\cos\phi\pm\sin\alpha\sin\phi&-\cos\alpha\sin\phi\pm\sin\alpha\cos\theta\cos\phi\\
			0&\sin\theta\sin\phi&\cos\alpha\cos\theta\sin\phi\mp\sin\alpha\cos\phi&\cos\alpha\cos\phi\pm\sin\alpha\cos\theta\sin\phi\\
			0&\cos\theta&-\cos\alpha\sin\theta&\mp\sin\alpha\sin\theta
		\end{array}
		\right),
		\label{r3}
	\end{equation}
\end{widetext}
where we have restored the first row and the first column related to time.
Eq.\eqref{r3} is actually a spatial rotation with the intrinsic Euler angles
$\psi_r=\pm\alpha-\frac\pi2$, $\psi_\theta=\theta-\frac\pi2$ and $\psi_\phi=\phi$
around the original axes $\hat r$, $\hat\theta$ and $\hat\phi$, repectively.

Taking this rotation into the Eq.\eqref{lambda0},
we then have a $\Lambda\in\text{SO}(1,3)$ that transforms the diagonal tetrad \eqref{diage} into a new one.
The spin connection can be obtained using Eq.\eqref{omega0}:
\begin{equation}
	\begin{split}
		\omega^{\hat0}_{\:\hat1}&=\omega^{\hat1}_{\:\hat0}=\frac{\xi'}{2\xi\sqrt{1-\xi}}\diff r,\\
		\omega^{\hat2}_{\:\hat3}&=-\omega^{\hat3}_{\:\hat2}=\pm\frac{\xi\zeta'-\zeta\xi'}{2\zeta\sqrt{\xi\zeta-\xi^2}}\diff r-\cos\theta\diff\phi,\\
		\omega^{\hat0}_{\:\hat2}&=\omega^{\hat2}_{\:\hat0}=-\sqrt{\frac{1-\xi}\zeta}\diff\theta\pm\sqrt{\frac{(1-\xi)(\zeta-\xi)}{\xi\zeta}}\sin\theta\diff\phi,\\
		\omega^{\hat0}_{\:\hat3}&=\omega^{\hat3}_{\:\hat0}=\mp\sqrt{\frac{(1-\xi)(\zeta-\xi)}{\xi\zeta}}\diff\theta-\sqrt{\frac{1-\xi}\zeta}\sin\theta\diff\phi,\\
		\omega^{\hat1}_{\:\hat2}&=-\omega^{\hat2}_{\:\hat1}=-\frac1{\sqrt\zeta}\diff\theta\pm\sqrt{\frac{\zeta-\xi}{\xi\zeta}}\sin\theta\diff\phi,\\
		\omega^{\hat1}_{\:\hat3}&=-\omega^{\hat3}_{\:\hat1}=\mp\sqrt{\frac{\zeta-\xi}{\xi\zeta}}\diff\theta-\frac1{\sqrt\zeta}\sin\theta\diff\phi,\\
	\end{split}
	\label{omeganew}
\end{equation}
where we have substituted that $\cos\alpha=\sqrt{\xi/\zeta}$.
And the new tetrad can be obtained by performing the transform $\Lambda$ on the diagonal tetrad \eqref{diage}:
\begin{equation}
	\begin{split}
		\mathring h^{\hat0}=&\diff t-\sqrt{\frac\zeta\xi-\zeta}\diff r,\\
		\mathring h^{\hat1}=&-\sqrt{1-\xi}\sin\theta\cos\phi\diff t+\sqrt{\frac\zeta\xi}\sin\theta\cos\phi\diff r\\
		&+\left( \sqrt{\frac\xi\zeta}\cos\theta\cos\phi\pm\sqrt{1-\frac\xi\zeta}\sin\phi \right)r\diff\theta\\
		&-\left( \sqrt{\frac\xi\zeta}\sin\phi\mp\sqrt{1-\frac\xi\zeta}\cos\theta\cos\phi \right)r\sin\theta\diff\phi,\\
		\mathring h^{\hat2}=&-\sqrt{1-\xi}\sin\theta\sin\phi\diff t+\sqrt{\frac\zeta\xi}\sin\theta\sin\phi\diff r\\
		&+\left( \sqrt{\frac\xi\zeta}\cos\theta\sin\phi\mp\sqrt{1-\frac\xi\zeta}\cos\phi \right)r\diff\theta\\
		&+\left( \sqrt{\frac\xi\zeta}\cos\phi\pm\sqrt{1-\frac\xi\zeta}\cos\theta\sin\phi \right)r\sin\theta\diff\phi,\\
		\mathring h^{\hat3}=&-\sqrt{1-\xi}\cos\theta\diff t+\sqrt{\frac\zeta\xi}\cos\theta\diff r\\
		&-r\sqrt{\frac\xi\zeta}\sin\theta\diff\theta\mp r\sqrt{1-\frac\xi\zeta}\sin^2\theta\diff\phi.
	\end{split}
	\label{mringh}
\end{equation}

\section{Flat spacetime limit and some examples}
\label{flatlimit}
In flat limit $\xi=\zeta=1$, the tetrad \eqref{mringh} becomes
\begin{equation}
	\mathring h^a|_{\xi=\zeta=1}=\left( \diff t,\;\diff x_{\hat1},\;\diff x_{\hat2},\;\diff x_{\hat3} \right),
	\label{holo}
\end{equation}
namely holonomic, and the torsion 2-form $T^a=\diff\mathring h^a=\diff\diff x_a$ vanishes
since the components of $\mathring h^a|_{\xi=\zeta=1}$ are exact, which is expected for flat spacetime.
This means that the legs of the tetrad \eqref{holo} are aligned everywhere
and the spurious inertial effects are removed,
which makes Eq.\eqref{holo} ideal for the proper tetrad for flat spacetime.
In this case, the spin connection \eqref{omeganew} becomes
\begin{eqnarray}
	\omega^{\hat2}_{\;\hat1}=\diff\theta,\quad\omega^{\hat3}_{\;\hat1}=\sin\theta\diff\phi,\quad\omega^{\hat3}_{\;\hat2}=\cos\theta\diff\phi,
	\label{ksomega}
\end{eqnarray}
which is the same as the spin connection introduced in Ref.\cite{Krssak2016}.
Together with the non-flat diagonal tetrad \eqref{diage},
this spin connection gives a specific choice of proper tetrad introduced in Refs.\cite{Lucas2009,Krssak2016}:
\begin{equation}
	\begin{split}
		\tilde h^{\hat0}=&\sqrt\xi\diff t\\
			\tilde h^{\hat1}=&\sqrt\zeta\sin\theta\cos\phi\diff r+r\cos\theta\cos\phi\diff\theta+r\sin\phi\sin\theta\diff\phi\\
			\tilde h^{\hat2}=&\sqrt\zeta\sin\theta\sin\phi\diff r+r\cos\theta\sin\phi\diff\theta+r\cos\phi\sin\theta\diff\phi\\
			\tilde h^{\hat3}=&\sqrt\zeta\cos\theta\diff r-r\sin\theta\diff\theta.
	\end{split}
	\label{kstetrad}
\end{equation}
Eq.\eqref{kstetrad} also reduces to the holonomic \eqref{holo} in the flat limit $\xi=\zeta=1$.
That is, the new proper tetrad \eqref{mringh} coincides with the proper tetrad introduced in Refs.\cite{Lucas2009,Krssak2016} in flat limit.
In fact, utilizing the flat limit as the reference tetrad and solving
\begin{equation}
	T^a|_{\xi=\zeta=1}=\diff h^a+\omega^a_{\;b}\wedge h^b=0
	\label{reftetrad}
\end{equation}
for $\omega^a_{\;b}$ is exactly how Eq.\eqref{ksomega} is obtained\cite{Krssak2016,Krssak2018},
where $h^a$ is any given tetrad, and in the current case it is the diagonal one \eqref{diage}.
This procedure bears a simple but important facts that the torsion 2-form should ideally vanish in flat spacetime limit.
However, as is shown, the tetrad \eqref{kstetrad} is not the only choice of proper tetrad that can carry out this idea.
For non-flat spacetime \eqref{sssmetric},
spin connections with certain combinations of $\xi$ and $\zeta$ can also make it,
as long as they reduce to Eq.\eqref{ksomega} in the limit $\xi=\zeta=1$.
Or equivalently, certain tetrads are also suitable to be considered as the proper tetrad if they can reduce to Eq.\eqref{holo} in the flat limit.
Undoubtedly, our new tetrad \eqref{mringh} is one of them.

In the underlying static spherically symmetric case,
the new proper tetrad \eqref{mringh} will lead to the torsion scalar
\begin{equation}
	T=\frac2{r^2\zeta^2}\left( \zeta^2-\zeta+r\zeta' \right),
	\label{Tproper}
\end{equation}
whereas if one uses Eq.\eqref{kstetrad} as the proper tetrad\cite{Lucas2009,Krssak2016}, the torsion scalar will be
\begin{equation}
	T=-\frac{2\left( \sqrt\zeta-1 \right)\left( \xi-\sqrt\zeta\xi+r\xi' \right)}{r^2\zeta\xi}.
	\label{KSTgen}
\end{equation}
In certain teleparallel gravities like TEGR or $f(T)$ gravity,
the difference between Eqs.\eqref{Tproper} and \eqref{KSTgen} will lead to different behaviors of the field equations
in that Eq.\eqref{KSTgen} will encounter apparent singularities at the possibly existing horizon(s) where $\xi\rightarrow0$,
while Eq.\eqref{Tproper} will behave regularly.
Hence the new proper tetrad and the tetrad introduced in Refs.\cite{Lucas2009,Krssak2016}
describe physically different strong field situations.

For example, in TEGR, due to the equivalence to GR,
Schwarzschild vacuum $\xi=1/\zeta=1-\frac{r_\text s}r$ is surely one of the solutions.
In fact, different choices of proper tetrad can be utilized to obtained this solution in TEGR\cite{Krssak2018}.
In this case,
the tetrad in \cite{Lucas2009,Krssak2016} leads to
\begin{equation}
	T=-\frac2{r^2}\left( \sqrt{\frac{r-r_\text s}r}+\sqrt{\frac r{r-r_\text s}}-2 \right),
	\label{TKSsch}
\end{equation}
which is obviously divergent at the horizon $r=r_\text s$,
whereas the new proper tetrad constructed in this paper leads to $T=0$.

For general $f(T)$ theories with the action
\begin{equation}
	\mathcal S=-\frac1{16\pi G}\int ef(T)\diff^4x+\int e\mathcal L_\text m\diff^4x
	\label{ftaction}
\end{equation}
where $e=\det(e^a_{\:\mu})$ and $\mathcal L_\text m$ is the matter Lagrangian,
we consider a simple model $f(T)=T+\lambda T^2$ here for exemplification.
The new proper tetrad leads to the static spherically symmetric field equations
\begin{equation}
	\left( \zeta^2+r\zeta'-\zeta \right)\left[ \left( r^2+2\lambda\right)\zeta^2+2r\lambda\zeta'-2\lambda\zeta \right]=0,
	\label{T200}
\end{equation}
and
\begin{equation}
	\begin{split}
		&\xi\left[ r^2\left( \zeta^3-\zeta^4 \right)+2\lambda\left( 2\zeta^3-\zeta^4+r^2\zeta'^2-\zeta^2 \right) \right]\\
	&+r\xi'\left[ \zeta^3r^2+4\lambda\zeta\left( \zeta^2+r\zeta'-\zeta \right) \right]=0.
	\end{split}
	\label{T211}
\end{equation}
The solutions would be
\begin{equation}
	\xi=\frac1\zeta=1-\frac{r_s}r,\quad\text{or}\quad\xi=\frac1\zeta=1+\frac{r^2}{6\lambda}-\frac{r_d}r,
	\label{zetaT2}
\end{equation}
where $r_s,r_d$ are integral constants,
which means that the only solutions in $T+\lambda T^2$ model
are the Schwarzschild one and the Schwarzschild-(anti) de Sitter one.
On the other hand,
when taking the tetrad \eqref{kstetrad} as the proper tetrad,
it is known that the Schwarzschild vacuum is not a solution to the field equations if $f(T)$ is not a linear function of $T$\cite{Bohmer2011,Tamanini2012}.

In the cosmological scenario,
\begin{equation}
	\diff s^2=\diff t^2-\frac{a^2}{1-kr^2}\diff r^2-a^2r^2(\diff\theta^2+\sin^2\theta\diff\phi^2)
	\label{cosmetric}
\end{equation}
where $a(t)$ is the scale factor,
a ``good'' tetrad has been preferred\cite{FERRARO2011,Tamanini2012,Lin2017jcap} in pure-tetrad formalism.
The non-vanishing components of the spin connection associated with this preferred tetrad for $k=+1$ has also been listed in Ref.\cite{Hohmann2018}.
Since in this scenario the freely falling frame is the comoving one,
our construction of proper tetrad coincides with this preferred tetrad.

\section{Conclusion and discussions}
\label{discussions}
In the present paper,
we have constructed a new tetrad that can be considered as the proper tetrad for non-flat spacetime.
The new tetrad coincides with the tetrad proposed in Refs.\cite{Lucas2009,Krssak2016} in flat limit,
and hence can also realize the simple but important idea that the torsion 2-form should vanish in flat spacetime.
In cosmological scenario,
our construction of proper tetrad coincides with the preferred ``good'' tetrad\cite{FERRARO2011,Tamanini2012,Lin2017jcap} in pure-tetrad formalism.
In the strong field regime,
the new tetrad leads to different behaviors of the field equations in that
they will not encounter singularities at the possibly existing horizon(s),
thus can help with the search and analysis of the strong field solutions in teleparallel gravities.

With this new proper tetrad,
the Schwarzschild vacuum leads to a vanishing torsion scalar $T=0$.
Since in torsion-based gravities, the torsion scalar $T$ takes a position similar to that of the Ricci scalar $R$ in curvature-based gravities,
this result is an direct analogue of the vanishing $R$ in the Schwarzschild vacuum.
Due to the equivalence,
the Schwarzschild vacuum is no doubt one of the solutions in TEGR.
Although any tetrad including the diagonal one can be used to obtain this solution,
it is still necessary to consider the choice of proper tetrad in TEGR
for the regularization reason\cite{Lucas2009,Krssak2015}.
Simple calculation shows that our proper tetrad
will also successfully regularize the four-momentum for the Schwarzschild metric.

Furthermore, using the new proper tetrad,
we have also consider a simple model of $f(T)$ gravities with $f(T)=T+\lambda T^2$,
which has been previously considered in various cases\cite{DeBenedictis2016,Farrugia2016,Lin2017,Mai2017}.
Here we have proved that this model has only the Schwarzschild and Schwarschild-(anti) de Sitter solutions.
This suggests that with the new proper tetrad,
the TEGR solutions are also solutions to $f(T)$ as argued in Refs.\cite{Ferraro2011a,Tamanini2012},
and hence there is a greater chance to find physical application for $f(T)$ gravities.
This result may be extended to a more general $f(T)$ model.
It is known that in $f(R)$ gravities\cite{Psaltis2008,Bergliaffa2011},
setting $R=\text{const.}$ will lead to the absence of new solution other than the general relativistic ones.
Without this setting,
the possibility of having new solutions in fact lies in the higher order derivatives of the field equation.
Since $f(T)$ gravities feature in second order derivatives as in GR,
it will not be surprising that they have only the current solutions.
Nonetheless, at perturbation level,
it is still possible to tell apart the $f(T)$ solutions from the GR ones as in $f(R)$ cases\cite{Barausse2008}.
A general $f(T)$ model and its perturbation will be considered in the future.

The study presented in the current paper shows that
there are different choices of tetrad that can be considered as the proper tetrad for teleparallel gravities.
Investigating this issue may hopefully pave the way to finding and analyzing the spherically symmetric solutions,
such as black holes, wormholes and other astrophysical stars in teleparallel gravities,
as well as to the construction of strong field solutions with other symmetry.
With the historic detection of gravitational waves,
it is important to be able to study the strong field behaviors of teleparallel gravities if there is any chance in testing them using this powerful tool.

\begin{acknowledgments}
This work is supported by the National Science Foundation of China under Grant Nos.~11847080 and ~10671128,  and the Key Project of Chinese Ministry of Education under Grant No.~211059.
\end{acknowledgments}

\bibliography{ref}
\bibliographystyle{apsrev}

\end{document}